\documentclass[aps, superscriptaddress,prl,amssymb, amsmath,twocolumn]{revtex4}
\usepackage{amsmath,latexsym,amssymb}
\usepackage[dvips]{graphicx}
\usepackage{amsmath}
\usepackage{latexsym}
\usepackage{amssymb}
\usepackage{bm}

\newcommand{\be}{\begin{eqnarray}}
\newcommand{\ee}{\end{eqnarray}}

\newtheorem{thm}{Theorem}
\newtheorem{defin}{Definition}
\newtheorem{obs}{Observation}

\def\LHV{{LHV model}}
\def\source{{\it source}}
\def\g{{\bf g}}
\def\d{{\bf d}}

\def\ep{{\epsilon}}

\def\V{{\bf V}}
\def\v{{\bf v}}
\def\X{{\bf X}}

\def\Y{{\bf Y}}

\begin{document}

\title{When non i.i.d. information sources can be communicationally useful?}

\author{Marcin Paw{\l}owski}
\affiliation{Institute of Theoretical Physics and Astrophysics, University of Gda\'nsk, 80-952 Gda\'nsk, Poland}
\author{Karol Horodecki}
\affiliation{National Quantum Information Center of Gda\'nsk, 81-824 Sopot, Poland}
\affiliation{Institute of Informatics, University of Gda\'nsk, 80-952 Gda\'nsk, Poland}
\author{Pawe{\l} Horodecki}
\affiliation{National Quantum Information Center of Gda\'nsk, 81-824 Sopot, Poland}
\affiliation{Faculty of Applied Physics and Mathematics,
Gda{\'n}sk University of Technology, 80-952 Gda{\'n}sk, Poland}
\author{Ryszard Horodecki}
\affiliation{Institute of Theoretical Physics and Astrophysics, University of Gda\'nsk, 80-952 Gda\'nsk, Poland}
\affiliation{National Quantum Information Center of Gda\'nsk, 81-824 Sopot, Poland}


\begin{abstract}
Information sources used for communication purposes usually are assumed
to be i.i.d. type, especially as far as entanglement or nonlocality properties
are concerned. Here we proposed simple scheme for detection of nonlocality of non i.i.d.
sources based on the program idea which can be useful in communication tasks including 
quantum cryptography and communication complexity reduction.  The principle of the scheme is rather 
general and can be applied to other problems like detection of entanglement coming 
from non-i.i.d sources.
\end{abstract}

\maketitle

{\it Introduction .-}  The problem of ,,locality vs nonlocality'' was usually treated under the
assumption that the information source is one which emits signals that are independent from each other. Such sources can be described by a sequence of  identical and independently distributed (i.i.d.) random variables. However there may exist sources for which the assumption may not work for purely physical reasons.
The main idea of the paper is to propose a simple scheme for detection
of nonlocality via Bell inequalities (or detection of quantum entanglement in terms of witnesses) in (in general multipartite) systems which are non i.i.d, but have some type of classical (but not quantum) memory.
   Let us mention the first approach to non i.i.d. sources was proposed in Ref. \cite{Datta-Non-iid-06}
in terms of entropic quantities and then developed for entanglement measures \cite{Datta-Cost-07} and classical capacities \cite{Datta-Capacity-08} (see also Ref. \cite {Renner-non-iid-08}). Here however we are focused on {\it qualitative}
detection of quantum properties like nonlocality and entanglement. In particular the nonlocality
detection needs to be more rigorous.
   The key notion - in contrary to standard approach to nonlocality - is the program that
is independent from Alice and Bob settings (that are chosen randomly) and allow to subselect
the substrings of the ,,truly'' nonlocal (or ,,truly'' entangled ) data. Usually we shall
assume that the source has no prior information about Alice and Bob settings choices.
We show how this assumption can be relaxed in some cases.
 We discuss the result for bipartite case, however the scheme works also for multipartite versions.

{\it Scenario and the main idea.-}  In bipartite version the scenario we consider involves spatially separated Alice and Bob and a \source\ S. The source is not specified, apart from the assumptions, that for each $N$ it provides Alice and Bob a sequence of $N$ objects each of which both of them can measure, and it does not have access to their settings. The task of Alice and Bob is to check if the source can be turned into a one which does not 
admit an LHV model.

Let us remind that in particular the source need not lead to independent and identically distributed random variables, 
as assumed in usual Bell inequality tests. We are rather closer to the new paradigm that has been considered only recently in \cite{Massanes-Bell-07}, where the source can be totally empowered by the eavesdropper Eve. However we do not adopt the idea of symmetrization used there, since for certain patterns of memory the source which is quite useful, would be claimed to be useless, as admitting an LHV model. Exemplary source is the one which is 'local' and 'nonlocal' alternately, corresponding to e.g. an unstable device.

The idea is quite simple. If Alice and Bob observe, that their data do not violate given Bell
inequality, they are not yet lost. They can try to design a proper {\it program} such that after processing of their (classical) data, they will observe the violation. By {\it program} we mean any algorithm written for any (classical, quantum, hybrid or other) machine that filters some events i.e. instructs which outcomes of measurements should be taken into account, and which to be discarded. By proper, we mean that the {\it program} is independent from the {\it settings} (choices of measurements involved in Bell inequality test)
but in principle it may depend on some other data like some additional measurement results (as long as they
are settings-independent too) or even laboratory conditions, weather etc. This aspect is common with 
the original concept of hidden nonlocality \cite{Popescu,RHN-extensions}. The later is included in our model
as a special case, however in the present analysis we shall focus on programs dependent on Bell experiment events only, which is still vital since we deal with non i.i.d. sources.

 In other words, if Alice and Bob have access to a source $S$ which appears to be not directly useful, sometimes they can transform it into a useful one. It can be written symbolically as $S' =S+${\tt P} where {\tt P} is their program which 'filters' objects coming from $S$. 

To make the work with programs more easy, we first allow for Alice and Bob to transform their
data into a binary string, which is called $\g$. The more 1's in the string, the more probable it is to violate the Bell inequality. Hence the program {\tt P} should be chosen such that it instructs about as much of those entries of the string $\g$ which have 1, as it is possible, without being dependent from the settings in Bell experiment. 
A substring of $\g$ taken in positions indicated by {\tt P} forms a string called $\g$'. 
One may be also interested what happens if the program is somewhat correlated with the settings.
As we shall show if the string $\g$' is compressible enough, i.e. has low Kolmogorow complexity, and observed violation on $\g$' is high enough, we still may exclude LHV in original source.

In what follows, the letter in bold such as $\g$ will denote strings. The letters with indices such as $g_i$ will denote particular values of the $i$-th entry of the string. A bold capital letter will denote random variable.


{\it Sequence \g - the idea of binary G function.-} Consider an arbitrary linear two-partite Bell inequality, for
correlations of local outcomes observed at measurement stations of
Alice (A) and Bob (B) in the following form which avoids absolute values:
\be
\label{bell}
\sum_{a,b,x,y} \alpha(x,y,a,b) P(a,b|x,y) \leq R.
\ee
Here $x$ and $y$ stand for the measurement settings chosen by Alice and Bob respectively, and $a$ and $b$ for the outcomes of their measurements, $P(a,b|x,y)$ represents conditional probabilities and
$R$ is the local realistic bound. Note that any linear Bell inequality can be brought to the above form
written in such a form with  $\alpha(x,y,a,b)\geq 0 $ since
each inequality which has some negative $\alpha$'s can be rewritten by substituting
probabilities which stand by negative $\alpha$'s by unity minus the
probability of the opposite events.
With further notation
\be
\alpha(x,y,a,b)=P(x,y)C(x,y,a,b)G(a,b,x,y)
\label{alfa}
\ee
where $C(a,b,x,y)$ is any real, positive function and $G(a,b,x,y)\in \{0,1\}$ (\ref{bell}) becomes
\be \label{bellN}
\sum_{x,y} \sum_{a,b}  C(a,b,x,y)G(a,b,x,y)P(a,b,x,y) \leq R
\ee

When one wants to experimentally find the value of the LHS of (\ref{bellN}) one does $N$ runs of the experiment and approximates
\be
P(a,b,x,y)\approx\frac{\sum_{i=1}^N\delta_{a_i,a}\delta_{b_i,b}\delta_{x_i,x}\delta_{y_i,y}}{N}
\ee
where variables with index $i$ correspond to the values obtained in the $i$-th run of the experiment. When we plug this to (\ref{bellN})
we get
\be
\label{ubells}
\sum_{i=1}^N C_iG_i \leq RN
\ee
with abbreviations $C_i=C(a_i,b_i,x_i,y_i)$ and $G_i=G(a_i,b_i,x_i,y_i)$.
The form (\ref{ubells}) is very useful since it is clearly seen which run of the experiment raises the value of the LHS (the ones with $G=1$) so if one would like to make some 
postselection to increase the LHS one needs to choose these runs. The string of $N$ values of $G$ will be denoted as \g : \g $= (G_1,G_2,...,G_N)$.


{\it Choosing a substring with nonlocal properties - the idea of the \d\ string .-}
Let us assume that the strings {\bf c}=$(C_1,...,C_N)$ and \g=$(G_1,...,G_N)$ be such that they do not violate (\ref{ubells}) but there exists a subset of indexes $i\in {\cal I}$ such that
\be \label{V}
\sum_{i\in {\cal I}} C_iG_i \leq RN'
\ee
is violated, where $N'$ is the cardinality of the set $I$. The full description of ${\cal I}$ can be done with the help of the string
${\bf d} \in \{ 0,1\}^{N}$ of length $N$, where $[{\bf d}]_i=1$ if $i\in {\cal I}$ and $[{\bf d}]_i=0$ if $i \notin {\cal I}$. In what follows, string  will be the output of program {\tt P}. We will need also notation for substring of $\g$ indicated by ${\bf d}$ which will be denoted as $\g ' =(G_{\d(1)},...,G_{\d(N')})$.


{\it  The idea of the main result .-} From the construction of inequality (\ref{ubells}) it is clear, that if the source admits \LHV, then the string $\g$ satisfies this inequality. However, if $\g$ satisfies the inequality, the source that led to $\g$ {\it may still violate} the LHV model. We will provide now sufficient condition for this violation. Namely, the latter takes place if there is a binary string $\bf d$ such that substring $\g$' of $\g$ taken at positions imposed by $\bf d$ violates inequality (\ref{ubells}) and $\bf d$ 
can be reproduced by a {\it program} which is {\it independent from the settings}.
Of course, this fact should hold repetitively, that is Alice and Bob in order to test the source need to repeat the experiment $k$ times for sequence of length $N$. The number $k$ need not be large so that the probability that they can safely claim to see the source without LHV model is close to 1.


{\it The programs .-} In practice of laboratory, by a program we will mean any sequence of instructions written in any language on a personal computer, a hybrid classical-quantum computer, or any other machine. What is only important is that the program instruct (each $N$) how to transform an input string $\g$ of length $N$ into output $N'$ length string $\g$' formed from the subset of indices $1$ through $N$ (one can think about the program as a transition function of a deterministic Turing Machine) \cite{Papadimitriou}. We will actually not need its particular form, but can consider it indeed as collection of some functions transforming input into output. Thus, we adopt the following definition:

{\defin A program {\tt P} is a sequence of algorithms with their inputs 
given by $N$-bit sequences $\g$ and the outputs formed by binary $N$-bit sequences 
${\bf d}$.

 A program is independent from a random variable $\V$ if for each $N$ the
random variable {\bf P} of its output (induced by random variable of an input) 
is independent from $\V$ i.e.
\be
\forall_{{\bf w}_N^{(out)}}\, Pr(\V = \v | {\bf P} = {\bf w}_N^{(out)}) = Pr(\V = \v).
\ee
where ${\bf w}_N^{(out)}$ ranges the set of outputs of the program {\tt P} of length $N$ respectively.
\label{def:prog_indep}
}

There are two extreme examples of the programs, which clarify the idea of this (in)dependence defined above, as describe below.

{\it A negative Example}.-  Consider the following program:

{\tt "for an input containing length-N $\g$ string
output $\bf d=\g$.".
}
This program returns the string ${\bf d}$ which describes the substring $\g$'
with exactly all those values of $\g$ where function G indicates correlations in the experiment. 
It is therefore highly {\it dependent} from its input. Indeed from the output one can 
reproduce {\it the whole} input string $\g$.
Such a program is cheating from our point of view, as its output can not be taken into account
in context of violation of the Bell inequality: a proper post-selection can always lead to its 'violation'.

In what follows we will first focus on programs which are independent from the variable of settings. Let us note however, that if the input string is not correlated with the settings, the program (if it is short) can not be too much correlated with the settings. We develop this case in the reminder of this paper.

{\it A positive Example}.-
For positive case, consider a family of constant programs $\tt P$
which output the same string independently on their inputs.
It describes what we shall call the {\it simple program}:

{\tt "for input containing any N-length string ${\bf d}$ output 
the fixed binary string $\bf d$".}

We have then the following observation, the proof of which is obvious:
{\obs Simple program acting on input $\g$ is independent from the settings (\X,\Y).
}


{\it  Testing procedure and the main result.-} Alice and Bob will perform
certain testing procedure. If the procedure passes the test, the source with a high probability does not admit the LHV model. The parameter $\ep>0$ will capture accuracy of this test.
The procedure will be just sampling performed on the blocks of signals taken from the source. A good event will be if a given block of length $N$ violates inequality (\ref{ubells}) by a constant $r \geq r_0$ with $r_0 >0$ describing the threshold of violation (i.e. one expects $(R+r_{0})$ on the LHS of (\ref{ubells})). They will choose randomly $k$ out of $K$ blocks. If the amount of good events is satisfactory they accept the source as characterized by the program {\tt P}$_0$, and abort the protocol otherwise. By satisfactory we mean that the rest of untested blocks (not all of which need to violate (\ref{ubells})) still violate the inequality when $P_0$ applied to them.

Formally, Alice and Bob can proceed few times according to the following procedure: 

1. Produce binary string $\g$ of length $N$.

2. Choose a program {\tt P}$_0$ which does not depend on settings (for usual Bell inequalities approach, {\tt P}$_0$ is just an identity permutation)

3. Check if 
the output of {\tt P}$_0$ on $\g$ indicates  $N'$ length substring $\g$' of $\g$ that violates inequality (\ref{ubells}) by $r \geq r_0>0$, and abort otherwise.

4. Take $N_{total} =N\times K$ more results from the source. Choose randomly $k$ blocks of length $N$ each. For each of them perform step 3 and count the 'yes' answers.

5. If the number  $k_{good}$ of 'yes' answers satisfies:
\be
{k_{good}\over k} \geq {R\over R + r_0} + \ep,
\label{k-good}
\ee
accept, and reject otherwise.


Let us check that the condition given above guarantees violation of the Bell inequalities
by the untested blocks when {\tt P}$_0$ applied to them. 
The number $K_{good}$ of good events in remaining (untested) part of string of blocks
can be estimated by sampling lemma (\cite{Debbie}) as
$K_{good} \geq ({k_{good}\over k}-\ep)(K - k)$.
By linearity of (\ref{ubells}) we can add $K-k$ violation parameters obtained for
each block processed by {\tt P}$_0$. For the $K-k - K_{good}$ blocks we put the worst case,
that there is not only no violation but just LHS of (\ref{ubells}) is zero.
The violation of the untested string is then lower bounded by
$K_{good}\times (R+r_0)N'$.
Hence, by the above inequality, we obtain that the total violation of a string of length $N'\times (K-k)$ (the output of $K-k$ times processed blocks by {\tt P}$_{0}$) has the LHS of the inequality bounded from below by
\be
({k_{good}\over k} -\ep)(K-k)(R+r_0)N'
\ee
To have violation of (\ref{ubells}) for a string of length $K-k$, we need
the above value to be not less than $(K-k)\times N'\times R$ which 
gives the desired bound (\ref{k-good}). 

What we aim to show is that we deal here with 'true' violation, i.e. that the program does 
not create itself the violating data, but rather extracts the violating string.

We are ready to provide the main result of this paper. 


{\thm If the source $S$ passes the testing procedure that involves program $P_0$, the source $S$ which has access to program $P_0$ does not admit the LHV model with a high probability.
\label{thm:Kol_bound}
}

{\it Proof}.-
By the very assumption about sources, $S$ is independent from the variable of settings ($X,Y$).
Since by point $1$ of testing procedure $P_0$ is also independent from settings, we can safely
claim, that $S$ with access to program $P_0$ is a valid source, where by access we mean, that the output of $S$ is filtered by $P_0$ (only those signals which are idicating by string $\bf d$ are passed). It is now sufficient to check,
that Alice and Bob observe Bell violation from their block-wise postprocessing. It is easy to
see that the testing procedure is independent from settings as well, and can be treated as 
a soubprocedure of $P_0$. By considerations below its definition (a consequence of points $3-5$), we obtain that source $S$ equipped with a program $P_0$ (or - equivalently - Alice and Bob with this program, having access to the source) violates the Bell inequality. The probability that the procedure succeeds approaches 1 is exponentially fast in the number of tested blocks $k$ of in testing procedure providing $k \in O(\sqrt{K})$.

{\it Simple example .-} An elementary example is a quantum source sending infinite sequence with even (odd) 
two-qubit system in maximally entangled $|\Psi_+\rangle$ (separable $(I-|\Psi_+\rangle\langle \Psi_+|)/3$)
state. Observer testing standard CHSH inequality will get averages of separable two-qubit Werner state
which will obviously obey the inequality. However the following program {\tt P}: {\tt "take 
 $\d$ with $[\d]_i = 0$ $([\d]_i = 1)$ for the index $i$ even (odd)"} will pass the above testing procedure with maximal violation of the inequality 
($R=2\sqrt{2}$) which immediately the presence of LHV in the source.


{\it Bounds on programs correlated with settings .-} Sometimes it may happen 
that the program {\tt P} is correlated to the variable describing the settings. 
This problem is related to the possible restriction on ,,free will''
of the observers which may lead to false violation of LHV, 
but still nontrivial LHV bounds may be derived (see \cite{Correlated}).
Here we shall derive the bounds in $G$-string type Bell inequality 
for programs {\tt P} assumed to be correlated with settings.
For simplicity we shall consider the Bell-CHSH variant, however the 
idea naturally extends to any Bell inequality for which the rate of the outcomes 
implying $G=1$ to all outcomes maximized over all setting choices and hidden 
variables is bounded away from 1. 

Assuming that there is some correlation between hidden variable
$\lambda$ and the settings we get for fixed $\lambda$  Bell inequality component:
\be
\sum_{x,y=0}^1 P(x,y|\lambda) P(a \oplus b = x y | x,y) \le B
\label{CHSH-L}
\ee
where now $B \in[\frac{3}{4},1]$ depending on correlations between settings and
$\lambda$, but {\it not} on the $\lambda$ itself is to be found.
Note that of course here still $\sum_{\lambda}P(x,y|\lambda)=P(x,y)=\frac{1}{4}$
since the source has no influence of observers choices. The best strategies
(LHV functions) for the source are those which make the sum in LHS of (\ref{CHSH-L})
as large as possible. t is not hard to see that the best source strategy to maximize LHS of (\ref{CHSH-L}) for given four component vector $P(x,y|\lambda)$ is the one which multiplies by $P(a \oplus b = x y | x,y)=0$
the least component of the vector. This immediately gives the bound
\be
B=1-r,\ \ r=\min_{(x,y);\lambda}\{P(x,y|\lambda)\}
\ee
Now for given $B$ or - equivalently - $r$ let us estimate the mutual information $I(\lambda:XY)=H(XY)-H(XY|\lambda)$ form below. Clearly $H(XY)=2$ while
for fixed $r$ $H(XY|\lambda)$  is maximized by the vector $P(xy|\lambda)$
with all the components other than $r$ equal to each other:
\be \nonumber
H(XY|\lambda)\leq -r\log r-(1-r)\log\frac{1-r}{3}=f(r)
\ee
which gives
$I(\lambda: XY)\geq 2-f(r)$.
It means that if the local realistic bound is $B=1-r$ then the source
has at least $2-f(r)$ information about the settings.
On the other hand, since $f$ is monotonously increasing,
if the source has $I$ information about the settings then the local realistic bound is at most
\be
B\leq 1- f^{-1}(2-I) \equiv B(I)
\ee
Putting this bound into (\ref{CHSH-L}) and averaging it over $\lambda$'s
gives the new inequality involving {\it via} information $I$ the correlations
between the hidden parameter and the settings:
\be
\sum_{x,y=0}^1 P(x,y) P(a \oplus b = x y | x,y) \le  B(I)
\label{CHSH1}
\ee
with the critical amount of correlation information the source can have
(ie. the one reaching the quantum bound)
$I_{crit}=2-f(1-P_{QS})=2-f(\frac{1}{2}-\frac{1}{2\sqrt{2}}) \approx 0.046 $.

Note that the parameter $\lambda$ that saturates the latter bound 
can be interpreted as an informational ,,content'' (or ,,capacity'') 
of correlations contained in the source that correspond to quantum behavior.
If applied to the quantum state $\Psi$ it can be regarded as 
classical hidden simulation of correlation properties of the state
with respect to the considered Bell experiment. The role of this parameter 
(or - more precisely - equivalence class of such parameters) and its possible 
optimization over set of more experiments will be considered elsewhere.

For the source with fixed $I$ note that any program {\tt P}
producing the string of length $N'$
modifies the inequality (\ref{CHSH1}) leading for large $N'$ to:
\be
\sum_{i\in {\cal I}} C_i G_i \leq B(I) N'
\ee
Now consider the $N$ length binary string $\bf d$ which is 
binary characteristic function of substring $\g$' of $\g$ ie. 
it has zeros on the positions which were rejected by the program
and ones on those positions which carry those elements of  $\g$ that were 
chosen to form $\g$'. Suppose that Kolmogorow complexity of 
$\bf d$ is M.
It is clear that $\bf d$ cannot contain more than $M$ information about the settings. 
However this is the only information that is ,,moved back'' to the source 
in the proof of the  Theorem 1.
Since the settings form string of i.i.d., the information the  $\bf d$
holds about each individual settings follows
\be
\sum_{i\in {\cal I}} I_i \leq M
\ee
where $I_i$ is the information $\bf d$ holds about the settings in $i$-th round of experiment. 
The use of the program {\tt P} for the string from local realistic source is equivalent to another 
local realistic source which has this program built in. But even though it does not
meet the freedom of choice requirement it still leads to local realistic bound:
\be
\sum_{i\in {\cal I}} C_i G_i \leq \sum_{i\in {\cal I}}B(I_i)
\ee
Finally note that since $B(I)$ is a concave function one has
$\sum_{i\in {\cal I}} B(I_i)\leq B(\frac{M}{N'})N'$.
giving another inequality which is of our central interest:
\be
\sum_{i\in {\cal I}}C_i G_i \leq \Big(1-f^{-1}(2-\frac{M}{N'})\Big)N'
\ee
It is noteworthy that this gives a bound on CHSH regardless of the program {\tt P}.
So there are some procedures which lead to the bounds that even QM cannot 
violate. Our approach easily gives these bounds.


{\it Conclusions .-} We have provided natural method for extracting nonlocality properties
form the sources  that are non i.i.d.. If there exists some even long term memory in the
source the nonlocality (or entanglement) can be detected even if the standard tests fail.
It is interesting that, as we have shown, in the case of nonlocality it is possible to relax the condition
about no prior source knowledge about observers settings. Possible trade off between that knowledge and the power to perform communication tasks like communication complexity reduction or cryptography is an interesting subject for further research.
One can also consider programs with probabilistic algorithms built-in,  
especially in context of sources with Marcovian memory. These interesting
refinements of our scheme will be considered elsewhere. Note that the present scheme 
is quite general and may be applied to multipartite sources and Bell inequalities. 
Its principles may be also used for entanglement detection
and its applications like distillation and cryptography.
 

{\it Acknowledgements .-} We thank A. Grudka, W. Laskowski and M. Horodecki for discussion. 
K. H. thanks S. Pironio for indicating literature on similar topic. 
R. H. thanks C. Mora and M. Piani for discussion. 
The work was supported by UE SCALA project and by the LFPPI Network.


\end{document}